\documentstyle[twoside,fleqn,espcrc2]{article}
 
\newcommand{\bee} {\begin{equation}}
\newcommand{\ene} {\end{equation}}
\newcommand{\bea} {\begin{array}}
\newcommand{\ena} {\end{array}}
\newcommand{\beqa} {\begin{eqnarray}}
\newcommand{\enqa} {\end{eqnarray}}
\newcommand{\lapproxeq}{\lower .7ex\hbox{$\;\stackrel{\textstyle <}{\sim}\;$}}
\def\al{\alpha}

\def\tiz{SU(4)_{PS}\otimes SU(2)_L\otimes SU(2)_R}
\def\lu{SU(3)_c\otimes SU(2)_L\otimes SU(2)_R\otimes U(1)_{B-L}}
\def\sm{SU(3)_c\otimes SU(2)_L\otimes U(1)_Y}

\def\ul{SU(3)_c\otimes U(1)_Q}
\def\rw{\rightarrow}
\def\lrw{\longrightarrow}
\def\t{\tau_{p\rw e^+\pi^0}}

\newcommand{\AmS}{{\protect\the\textfont2
  A\kern-.1667em\lower.5ex\hbox{M}\kern-.125emS}}
 
\title{Cosmological implications of a class of $SO(10)$ models}
 
\author{
  G. Mangano\address{Dipartimento di Scienze Fisiche, Mostra 
    d'Oltremare Pad. 19, 80125 Napoli, Italia. \\ Istituto Nazionale di 
    Fisica Nucleare, Mostra d'Oltremare Pad. 20, 80125 Napoli, Italia.}
  and L. Rosa$^a$
       }
\begin{document}
 
\begin{abstract}
The cosmological implications of a class of $SO(10)$ models are discussed.
In particular we show how a good prediction for neutrino masses is obtained
in order to fit with the MSW mechanism to explain the solar neutrino flux
deficit and with the predicted amount of the dark matter {\it hot} component. 
A possible scenario for baryogenesis is also considered.
\end{abstract}
 
\maketitle
 
\section{INTRODUCTION}

It is well known that the electroweak standard model is now established
with incredible high precision. The dream of many physicists about
a higher degree of unification, realized by assuming a larger gauge
group at higher energies, cannot be tested by using the usual
{\it accelerator} approach. The typical signature of GUT theories, the
baryonic matter instability, has not been found so far but the lower limit on
the proton lifetime set the typical unification scale at $10^{15}-10^{16}
GeV$, which is incredibly high with respect even to next generation of
earth experiments. However, the big progress in astronomical observations
gives a chance to use the {\it history} of the universe as a natural device
to look for some evidence for an era, in the early stage of the universe,
governed by a larger symmetry. In this respect it is worth to study all 
cosmological implications of GUT models at high scales (inflation, 
topological defects, baryogenesis) and low scales (neutrino masses).\\
In this paper we will describe the results of a research on four particular 
$SO(10)$ symmetry breaking patterns. By studying the Higgs potential of the 
model, and using the Renormalization Group Equations, it was possible
to deduce \cite{lone} the values of the two physical 
scales of the theory: the unification scale, $M_X$, at which $SO(10)$ 
breaks to an intermediate group $G$, 
and the intermediate scale, $M_R$, at which the intermediate group is 
broken down to the standard group, $SU(3)_c \otimes SU(2)_L 
\otimes U(1)_Y \equiv SG$. Therefore, a typical $SO(10)$ 
breaking chain is given by

{\small
\bee
SO(10)\buildrel M_X\over\lrw G\buildrel M_R\over\lrw SG \buildrel M_Z\over
\lrw\ul. \label{eq:ssb}
\ene
}
\noindent $M_X$ is connected with the masses of the lepto-quarks which 
mediate proton decay, and, in particular, the present experimental lower limit 
on proton decay, $\t\geq 9\cdot10^{32}~years$ \cite{padb}, corresponds 
\cite{oliv} to the following lower limit on $M_X$:
\bee
M_X\geq 3.2\cdot10^{15}~GeV. \label{eq:mxinf}
\ene
Via the see-saw mechanism, $M_R$ is instead 
related to the masses of the (almost) left-handed neutrinos. 

\section{A CLASS OF $SO(10)$ MODELS}

It is quite a general result that the breaking of $SO(10)$ down to
the standard model group should proceed via an intermediate symmetry 
stage. In order to identify the possible directions for spontaneous 
symmetry breaking, one has to classify the components, in the smallest
irreducible representations (IR) of $SO(10)$, that are invariant under SG.

From Table 1 it is seen that for all IR, but the 144, the little group of 
the SG-singlet is larger than SG.

{\footnotesize
\begin{table}[tbh]
\setlength{\tabcolsep}{0.05pc}
\newlength{\ddigitwidth} \settowidth{\ddigitwidth}{\rm 0}
\caption{\baselineskip=11pt
{\small Classification of the Higgs $SG$-invariant components in 
the lower irreducible representation of $SO(10)$. $SG$ is the standard 
group while $D$ is the left-right discrete symmetry [5].}
}
\label{tab:class}
\begin{tabular}{@{}rcl}
\hline
                 \multicolumn{1}{c}{IR}             &
                 \multicolumn{1}{c}{$SG$}          &
                 \multicolumn{1}{c}{Symmetry}          \\
           &     \multicolumn{1}{c}{-singlet}       &  \\
\hline
16   &  $\hat{\chi}$    &  $SU(5)$                                          \\
45   &  $\hat{\al_1}$   &  $\lu$                                            \\
     &                  &  $\times D$                                       \\
45   &  $\hat{\al_2}$   &  $SU(4)_{PS}\otimes SU(2)_L\otimes U(1)_{T_{3R}}$ \\
54   &  $\hat{\sigma}$  &  $\tiz\times D$                                   \\
126  &  $\hat{\psi}$    &  $SU(5)$                                          \\
144  &  $\hat{\omega}$  &  $\sm$                                            \\
210  &  $\hat{\phi_1}$  &  $\lu$                                            \\
     &                  &  $\times D$                                       \\
210  &  $\hat{\phi_2}$  &  $\tiz$                                           \\
210  &  $\hat{\phi_3}$  &  $SU(3)_c\otimes SU(2)_L\otimes U(1)_{T_{3R}}\otimes 
U(1)_{B-L}$                                                                 \\
\hline
\end{tabular}
\end{table}
}

The use of 16, 126 and 144 representations would lead to the result that,
as in the minimal $SU(5)$ model, the three SG running coupling constant
would not meet at one point \cite{amal}.
Considering the 45 representation, one can 
show instead that the only non-trivial positive definite invariant with 
degree $\leq 4$ 
(as necessary in order to have a renormalizable potential) that one 
can build has its minimum in the {\small $SU(5) \otimes U(1)$}-invariant 
and its maximum in the {\small $SO(8) \otimes SO(2)$}-invariant directions
\cite{ext45}, so that it is not possible to construct a Higgs potential 
with minimum along the $\hat{\al_1}$ or $\hat{\al_2}$ directions.
Moreover, the $\hat{\phi_3}$ component in the 210 representation 
corresponds to a direction with neither {\small $SU(4)_{PS}$} nor {\small 
$SU(2)_R$} in the little group.

The previous considerations lead us to the following four choices for $G$

{\footnotesize
\[
  \bea{ll}
    SO(10) &                  \\ \\
    ~~(i)\buildrel\hat{\sigma}\over\lrw   & \tiz\times D  \\ \\
    ~(ii)\buildrel\hat{\phi_1}\over\lrw   & \lu\times D   \\ \\
    (iii)\buildrel\hat{\phi_2}\over\lrw   & \tiz          \\ \\
    (iv)\buildrel\hat{\phi_4}\over\lrw   & \lu,
  \ena
\]
}
\noindent where $\hat{\phi_4}= cos\theta\hat{\phi_1} + sin \theta\hat{\phi_2}$.
The spontaneous symmetry breaking to $SU(3)_c \otimes U(1)_Q$ is then 
realized, in all four cases, using the $\hat{\psi}$-component of 
a $126\oplus\overline{126}$ representation, and a combination 
of the {\small $\ul$}-invariant components of two 10's, in such a way to 
avoid the bad relation $m_t=m_b$ \cite{lone}.

The previous four possibilities have been studied in the 
Refs.\cite{lore}-\cite{luid}. We only report here for
brevity the results for the
case $(iv)$, which seems to be the more promising, 
for the mass scales corresponding to the two symmetry breaking
stage:
\bee
\bea{l}
M_R \leq 1.2\cdot 2.5^{0\pm 1}\cdot 10^{11}~GeV, \\ \\
M_X=1.9\cdot 2.0^{0\pm 1}\cdot 10^{16}~GeV.
\ena
\ene
\section{SOME COSMOLOGICAL IMPLICATIONS}
In the framework of the well known see-saw mechanism \cite{gera}, the upper 
limit for $M_R$ gives rise to the 
following inequalities for $m_{\nu_\tau}$ and $m_{\nu_\mu}$:
\bee
\bea{lll}
m_{\nu_{\tau}} &\geq& 34.1~ \frac{g_{2R}(M_R)}{f_3(M_R)}~eV  \\ \\
m_{\nu_{\mu}}  &\geq& 2.4\cdot10^{-3}~\frac{g_{2R}(M_R)}{f_2(M_R)}~eV,
\ena
\label{eq:mnu} 
\ene
where $g_{2R}$ and $f_i$ are the $SU(2)_R$ gauge coupling constant and the 
Yukawa coupling of the $126\oplus\overline{126}$ to the i-th family 
respectively, and we have used a value of the top quark mass $m_t = 176~ 
GeV$.
For reasonable values of $g_{2R}$ and $f_i$, Eqs.~(\ref{eq:mnu}) 
imply a substantial contribution of $\nu_\tau$ to the dark 
matter in the universe and a $m_{\nu_\mu}$ relevant for the MSW solution 
of the solar-neutrino problem \cite{msw}.

Another interesting prediction of this class of $SO(10)$ models is a 
dynamical explanation of the presently observed baryon asymmetry. Indeed, 
these models can satisfy the three necessary conditions stated by Sakharov 
\cite{sakh}: i) the $SO(10)$-gauge bosons mediate interactions which may lead 
to B-violations; moreover, at the scale $M_R$, 
 $B-L$ violating interactions may produce a net asymmetry in this quantum number
which cannot be washed out by the
baryon number violating electroweak anomalous effects, which are active at 
lower scales;
ii) at the intermediate scale, C and CP symmetry are 
broken\footnote{No $C$-odd quantum number asymmetries can be generated 
until C and CP symmetry remain unbroken, and this happens only at $M_R$ since 
the intermediate group has the same rank of $SO(10)$ \cite{mang}.}; iii) 
non-equilibrium conditions can be implemented if the masses of the Higgs 
particles satisfy certain conditions.

A scenario in which i), ii), and iii) are realized is discussed in 
Ref.\cite{mang}, in which the $SO(10)$ model with {\small $G=\lu$} is 
considered. A non-zero value of the asymmetry 
$\Delta (B-L)$ is produced at $T\sim
M_R$ by the $B-L$-violating decays $\tilde{\phi}\rw\tilde{\psi} f \nu_L^c$,
where $\tilde{\phi}$ are some Higgs multiplets of the 210 described in 
Ref.\cite{mang}, $\tilde{\psi}$ are the Higgs of the 126 which have mass of 
order $M_R$, and $f$ is a fermion. The knowledge of the mass spectrum of 
the Higgs scalars allowed to verify the 
possibility to have an overabundant population of $\tilde{\phi}$ at $M_R$, 
expressed by the inequalities $10^{12}GeV\leq m_{\tilde{\phi}}\leq 4\cdot 
10^{14}GeV$, where the first and second inequality correspond to the condition 
that the annihilation and decay processes respectively are "frozen out".
If no $B-L$-violating phenomena are active at 
lower temperatures, the stored $\Delta (B-L)$ is transformed in $\Delta B$ by 
the sphaleronic processes leading to a value for the photon to barion ratio
in agreement with the observed one $\eta_B = (4 \div 7) 10^{-10}$. 
Notice that the fact that an asymmetry in $B-L$
is produced, which cannot be washed out by the low-energy $B$ and $L$ anomalous
violating interactions, is quite crucial. In the usual $SO(10)$ baryogenesis
scenarios, in fact, it was assumed that B-violating interactions would provide
the observed value for the baryon to photon density ratio. The presence of
sphalerons, however, would completely erase any asymmetry produced in such a 
way.

\section{ACKNOWLEDGMENTS}
We are pleased to thank F. Buccella for useful discussions and comments.

\end{document}